# Uniaxial negative thermal expansion in a tetragonal $V_3S$-type superconductor $IrZr_3$

Yoshikazu Mizuguchi

*Department of Physics, Tokyo Metropolitan University, 1-1, Minami-osawa, Hachioji 192-0397, Tokyo, Japan.*

E-mail: mizugu@tmu.ac.jp

**Highlights:**
- Uniaxial negative thermal expansion was observed in $V_3S$-type $IrZr_3$.
- $IrZr_3$ is a superconductor with a transition temperature of 2.3 K.
- Uniaxial negative thermal expansion emerges in various transition-metal zirconides.

**Abstract:**

Recently, uniaxial negative thermal expansion (NTE) has been observed in transition-metal zirconides including tetragonal ($CuAl_2$-type) $CoZr_2$ and orthorhombic ($Re_3B$-type) $CoZr_3$. Here, we report the observation of uniaxial NTE in tetragonal ($V_3S$-type) $IrZr_3$. With increasing temperature, the lattice parameter for the $a$-axis and volume of $IrZr_3$ increase, but the lattice parameter for the $c$-axis decreases. $IrZr_3$ is a superconductor with a transition temperature of 2.3 K. The present result suggests that uniaxial NTE and related anomalous axis thermal expansion occur in various transition-metal zirconides, which will be important for designing superconducting device robust to temperature cycles using zero-thermal-expansion superconductors.

1. Introduction

Materials with characteristics of negative thermal expansion (NTE) or zero thermal expansion (ZTE) [1–5] have been utilized in various devices in which ultra-precision and small (or zero) displacements of position of components are required. Since ZTE compounds are rare as compared to NTE compounds, ZTE function is normally achieved by making composite materials of NTE and positive thermal expansion (PTE) compounds. Our study focuses on superconductors with anomalous axis thermal expansion. Thermal cycle has been a critical problem for superconducting materials and devices [6-8]. As well known, superconducting devices are used at very low temperatures and the fabrication process is basically done at higher temperatures. Hence, the affection of the thermal cycle to the device performance cannot be ignored. In addition, superconducting characteristics (superconducting states and performance) are generally sensitive to the presence of impurity and the stress to the grains. Therefore, discovery of ZTE superconductor is expected to improve the robustness of superconducting devices to thermal cycle. One of the ways to achieve ZTE in superconductor is the design of zero volume expansion in superconductors with uniaxial NTE. For example, in a tetragonal system, the NTE along the $c$-axis can compensate the PTE along the $a$-axis, which would result in the ZTE of the volume.

Recently, we reported uniaxial NTE in $CoZr_2$ and alloyed $TrZr_2$ ($Tr$: transition metal) with a tetragonal $CuAl_2$-type structure (space group: $I4/mcm$, #140) (see Fig. 1a) [9]. The observed NTE along the $c$-axis is relatively large among superconductors exhibiting uniaxial NTE [10-16]; the $c$-axis thermal expansion constant for $CoZr_2$ is $\alpha_c = -28$ $\mu K^{-1}$ [9]. Notably, volume ZTE has been seen in limited temperature range in $CoZr_2$ [9]. In $TrZr_2$, the thermal expansion is controllable by tuning $c/a$ ration of the tetragonal unit cell [17,18]. More recently, we reported the observation of uniaxial NTE in $CoZr_3$ with an orthorhombic $Re_3B$-type structure (space group: $Cmcm$, #63) (see Fig. 1b) [19]. Since uniaxial NTE has been observed in two different structural types, the uniaxial NTE in zirconides would be common characteristics. Therefore, to find out potential structural types where uniaxial NTE could be observed, we explored NTE in Ir-Zr compounds. Here, we show the uniaxial NTE in tetragonal $V_3S$-type (space group: $I-42m$, #121) $IrZr_3$. Although the observed NET in $IrZr_3$ is weaker than $CoZr_2$, the observation of uniaxial NTE in $IrZr_3$ provides us with a useful strategy to design new NET superconductors.

2. Material and methods

Polycrystalline samples (sample #1 and #2) of $IrZr_3$ were prepared by arc melting. Ir powder (99.9%, Kojundo Kagaku) and Zr plates (99.2%, Nilaco) were used for the syntheses. For sample #1, stoichiometric amount of $IrZr_3$ were melted. For sample #2, 1%-Zr-excess composition ($IrZr_{3.03}$) was used to compensate evaporation of Zr, but the quality of the obtained samples was comparable. Arc melting was performed in an Ar-filled chamber with two torches, and the samples were melted four times after turning over the sample. High-temperature X-ray diffraction (HTXRD) was performed on a Miniflex600 diffractometer (RIGAKU) equipped with a temperature-control device BTS-500 up to 573 K in vacuum. X-ray with Cu-K$\alpha$ radiation was used, and the diffraction data were collected using a semiconductor analyzer D/tex-Ultra (RIGAKU). Rietveld refinement was performed using RIETAN-FP [20], and the schematic images of crystal structure were drawn by VESTA [21]. Superconducting transition temperature ($T_c$) was measured by magnetic susceptibility on a superconducting quantum interference device (SQUID) magnetometer MPMS-3 (Quantum Design).

3. Results and discussion

As mentioned above, uniaxial NTE has been observed in CuAl$_2$-type CoZr$_2$ and $Tr$Zr$_2$ (Fig. 1a) [9] and Re$_3$B-type CoZr$_3$ (Fig. 1b) [19]. Here, we report on the observation of uniaxial NTE in V$_3$S-type IrZr$_3$ (Fig. 1c). Figure 2 shows the temperature dependence of magnetic susceptibility after zero-field cooling (ZFC) and field cooling (FC). A large negative susceptibility signals were observed below $T_c$ = 2.3 K, which is consistent with previous works [22,23].

Figure 3 shows the powder XRD pattern and the Rietveld refinement result for sample #1. Although small unknown peaks were included, almost all the peaks were explained with IrZr$_3$ (97%) and IrZr$_2$ (3%) with a cubic structure (space group: #227). Since the higher-angle peaks were well refined, the assumed structural model of V$_3$S-type IrZr$_3$ is reasonable. See table 1 for the refinement results. The refinement was performed with constant value of isotropic displacement parameter $U_{iso}$ = 0.127 Å$^2$ for both sites.

To evaluate temperature evolutions of lattice parameters, HTXRD were performed on two samples (#1 and #2). Figure 4 displays HTXRD patterns for sample #1 (see Supplementary material for HTXRD data for sample #2). To see the shifts of 00$l$ and $h$00 peaks, the 002 and 400 peaks were highlighted in Fig. 4b. The 400 peak continuously shifts to lower angles with increasing temperature, which indicates PTE along the $a$-axis. In contrast, the 002 peak slightly shifts to higher angles with increasing temperature, which suggests NTE along the $c$-axis. Figure 5 shows the temperature dependences of the obtained lattice parameters of $a$, $c$, and $V$. For both samples, PTE along the $a$-axis and NTE along the $c$-axis have been confirmed from the shift of lattice parameters. Two different samples exhibited comparable behavior of uniaxial NTE in IrZr$_3$. The estimated $\alpha_c$ for IrZr$_3$ is -4 μK$^{-1}$, which is clearly weaker than that observed in CoZr$_2$, and it seems to be saturated above 500 K. The results, however, propose that uniaxial NTE in $Tr$-Zr binary systems is not unique property of Co-Zr systems (CuAl$_2$-type CoZr$_2$ and Re$_3$B-type CoZr$_3$), and $Tr$-Zr binary systems with various structural types and wide-range of transition metals are potential materials for uniaxial NTE or anomalous axis thermal expansion. Since common origins are expected to present in various $Tr$-Zr binary compounds and related zirconides with uniaxial NTE, further investigations on crystal structure, electronic structure, and phonon characteristics of $Tr$-Zr systems are needed.

4. Conclusions

We synthesized polycrystalline samples of IrZr$_3$ with a V$_3$S-type tetragonal structure and investigated the superconducting and structural properties. Uniaxial NTE was observed along the *c*-axis in IrZr$_3$: with increasing temperature, the lattice parameters of *a* and *V* of IrZr$_3$ increase, but the lattice parameter of *c* decreases. From the susceptibility measurements, superconductivity with a $T_c$ of 2.3 K was confirmed in IrZr$_3$. The present result suggests that uniaxial NTE and related anomalous axis thermal expansion occur in various transition-metal zirconides and encourages further material development, characterization, and investigations on mechanisms in uniaxial NTE in *Tr*-Zr systems.

**Acknowledgements**

The author thanks O. Miura, H. Arima, A. Yamashita, and K. Mizuguchi for supports in experiments and discussion. This work was partly supported by the Tokyo Government Advanced Research (No.: H31-1), JSPS-KAKENHI (Nos.: 21H00151, 21K18834), and JST-ERATO (No.: JPMJER2201).

**Data statement**

The raw data reported in this paper can be provided by a reasonable request to the corresponding author.

Table 1. Structural parameter of IrZr$_3$ determined from Rietveld refinement (Fig. 1). Atomic coordinates are Ir: ($x$, 0, 0), Zr1: ($x$, $x$, 0.25), Zr2: ($x$, $x$, 0.25), and Zr3: ($x$, 0, 0.5).

| Sample | IrZr$_3$ (sample #1) |
|---|---|
| Space group | $I$-42$m$ (#121) |
| Temperature (K) | 296 |
| $a$ (Å) | 10.7875(3) |
| $c$ (Å) | 5.6573(2) |
| $V$ (Å$^3$) | 658.35(4) |
| $x$ (Ir) | 0.2908(3) |
| $x$ (Zr1) | 0.0907(4) |
| $x$ (Zr2) | 0.2932(3) |
| $x$ (Zr3) | 0.3515(3) |
| $R_{wp}$ | 9.4% |
| Ideal density (g/cm$^3$) | 9.400805 |

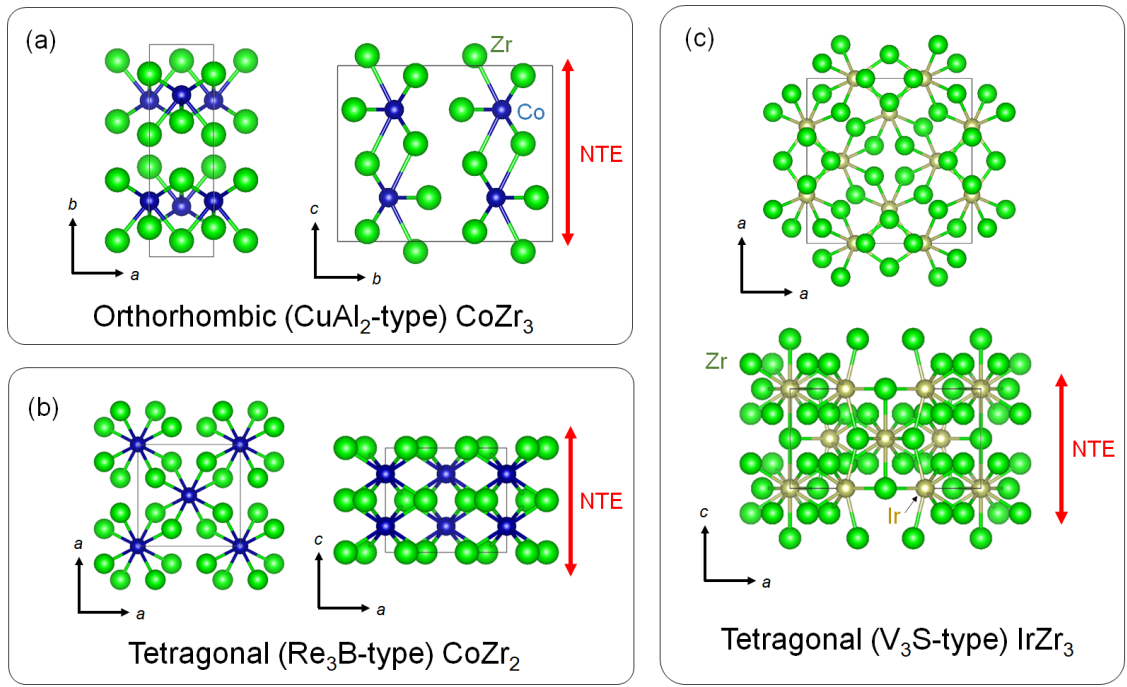

Fig. 1. Schematic images of crystal structure of (a) $CoZr_2$, (b) $CoZr_3$, and (c) $IrZr_3$. The squares indicate unit cells. NTE explains the direction of uniaxial negative thermal expansion.

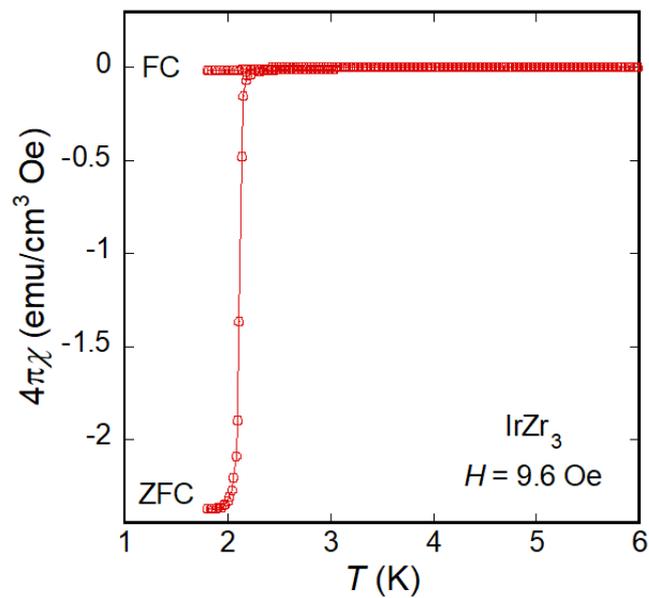

Fig. 2. Temperature dependence of susceptibility for $IrZr_3$ (sample #1). Both FC and ZFC data are shown.

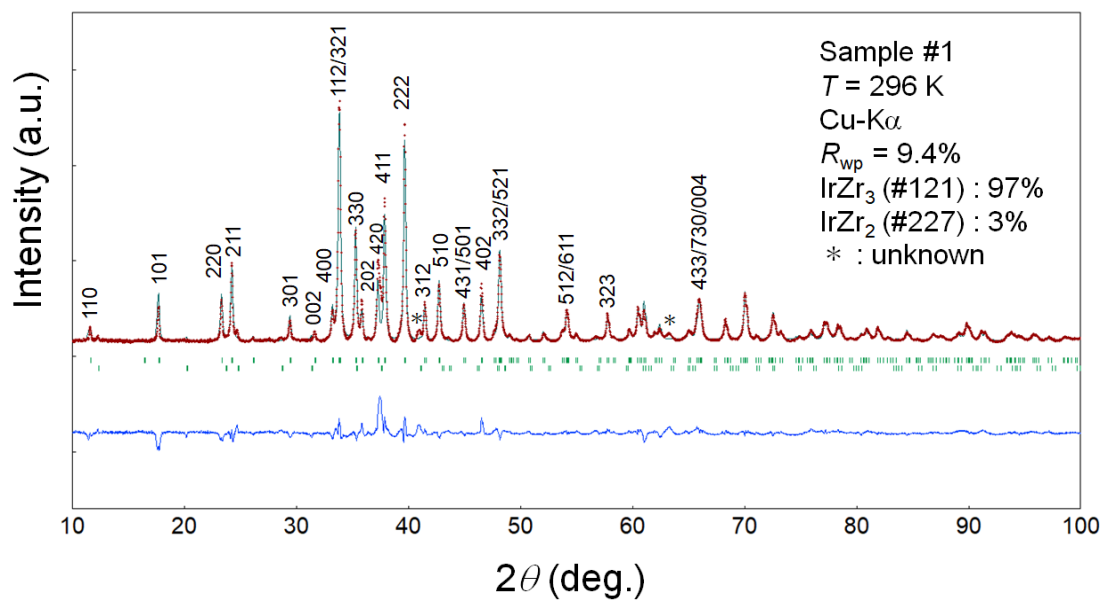

Fig. 3. Powder XRD pattern and Rietveld refinement result for IrZr$_3$ (sample #1). The peaks with asterisk were unidentified. Numbers in the figure are Miller indices.

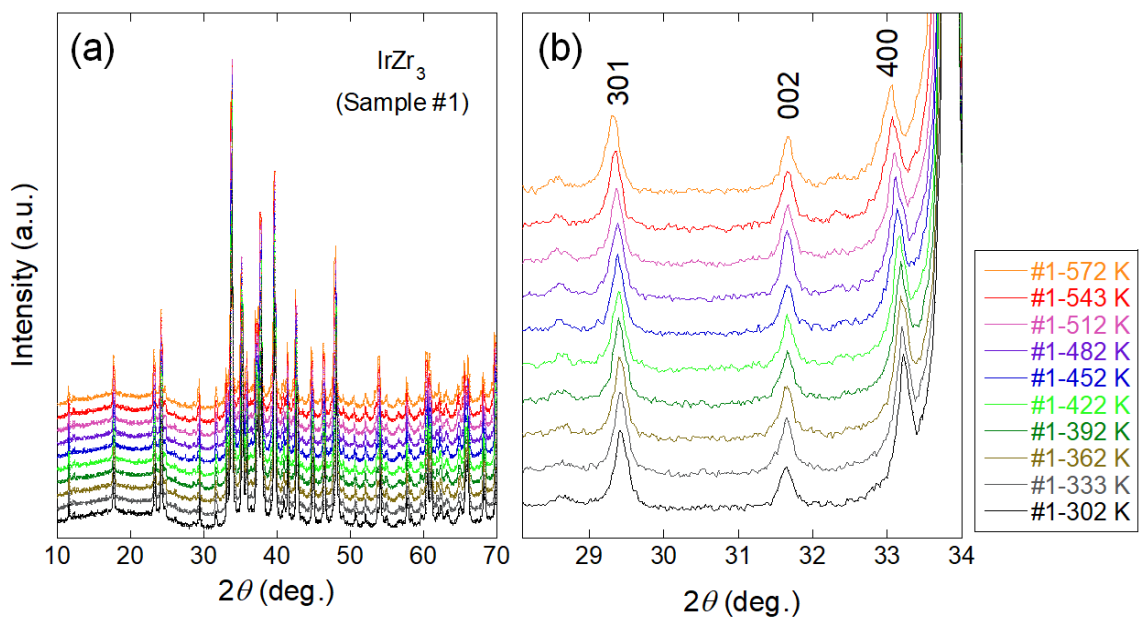

Fig. 4. (a) HTXRD patterns for IrZr$_3$ (sample #1). (b) Zoomed patterns near the 002 and 400 peaks.

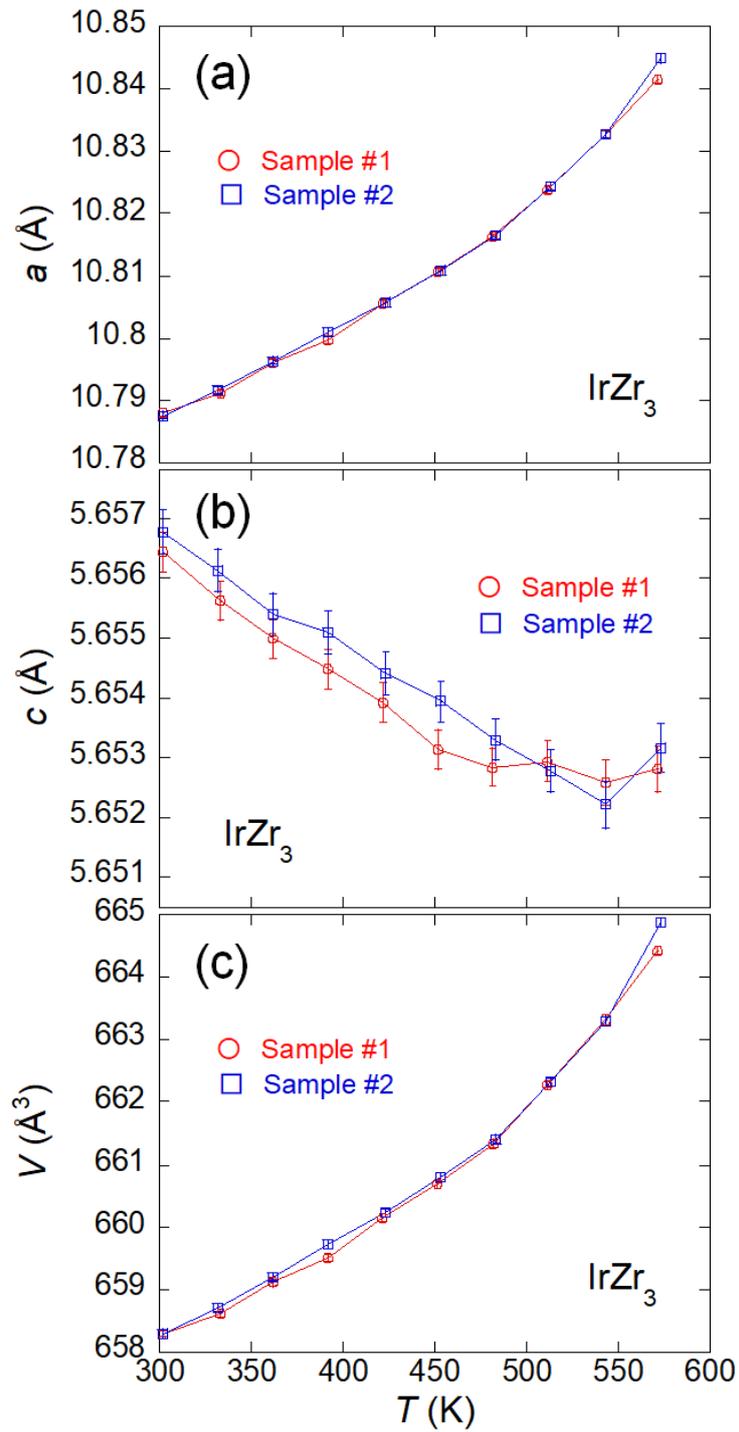

Fig. 5. Temperature dependences of lattice parameters of (a) $a$, (b) $c$, and (c) $V$ for the IrZr$_3$ samples (#1 and #2).

**Supplementary material**

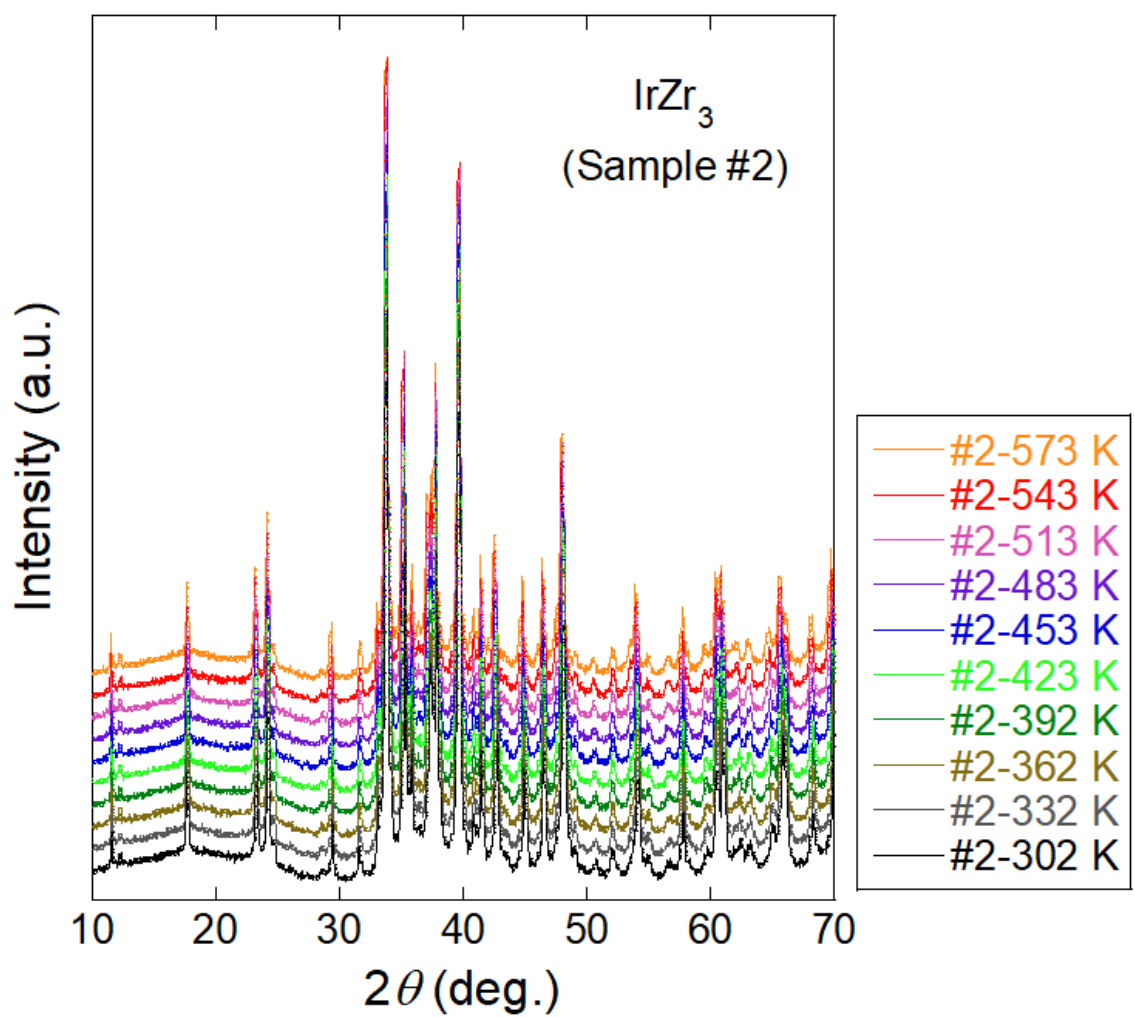

Fig. S1. HTXRD patterns for IrZr$_3$ (sample #2).